\documentclass[%
 preprint,
 amsmath,amssymb,
 aps,
]{revtex4-1}

\usepackage{graphicx}
\usepackage{dcolumn}
\usepackage{bm}
\usepackage[utf8]{inputenc}
\usepackage{hyperref}
\usepackage{color}

\begin{document}
\title {Enhancement of molecular coherent anti-Stokes Raman scattering with silicon nano-antennas}
  
\author{Shamsul Abedin$^1$, Yong Li$^2$, Abid Anjum Sifat$^3$, Khokan Roy$^2$, Eric O. Potma$^{2,3}$\email{epotma@uci.edu}}
\affiliation{$^2$Department of Chemical and Biomolecular Engineering, University of California, Irvine}
\affiliation{$^2$Department of Chemistry, University of California, Irvine}
\affiliation{$^3$Department of Electrical Engineering and Computer Science, University of California, Irvine}




\begin{abstract}
Surface-enhanced coherent anti-Stokes Raman scattering (SE-CARS) takes advantage of surface plasmon resonances supported on metallic nanostructures to amplify the coherent Raman response of target molecules. While these metallic antennas have found significant success in SE-CARS studies, photo-induced morphological changes to the nanoantenna under ultrafast excitation introduce significant hurdles in terms of stability and reproducilibty. These hurdles need to be overcome in order to establish SE-CARS as a reliable tool for rapid biomolecular sensing.  Here we address this challenge by performing molecular CARS measurements enhanced by nanoantennas made from high-index dielectric particles with more favorable thermal properties. We present the first experimental demonstration of enhanced molecular CARS signals observed at Si nano-antennas, which offer much improved thermal stability compared to their metallic counterparts.
\end{abstract}

\maketitle
\section{Introduction}
Surface-enhanced Raman scattering (SERS) is a popular tool for molecular sensing at ultralow analyte concentrations. In the conventional implementation of this technique, antennas composed of metallic nano-structures are used to efficiently capture and confine light to nanoscopic regions that contain the molecular species of interest~\cite{mccreery2005raman,fast2019coherent}. The molecule is efficiently driven by the confined optical fields in the electromagnetic ``hotspot'' and the nanoantenna then efficiently outcouples the induced near-field Raman polarization to far-field radiation, which is subsequently captured with a photodetector~\cite{Moskovits1985,willets2007localized,zhang2013chemical}. Ever since the inception of SERS in 1973~\cite{fleischmann1974raman}, numerous SERS enhancement structures have been devised and the technique has been successfully extended to capture Raman spectra from single molecules~\cite{nie1997probing,le2012single,Dieringer2007}. The mechanisms behind optical field localization and enhancement of the signal in SERS have been intensely studied and are generally well understood. While SERS allows enormous enhancement of the feeble Raman signal, acquisition rates beyond one to few tens of Hz are difficult to achieve in the single or few-molecule limit~\cite{Kneipp1998_nucleo,Chen2018}. DNA sequencing, rapid molecular screening and other high-throughput applications necessitate faster readout rates that SERS alone may not be able to provide.

Surface-enhanced analogues of coherent Raman scattering (CRS) methods are promising candidates to probe molecular vibrations at much faster rates compared to SERS. Surface-enhanced CRS was first demonstrated in 1979 where the coherent anti-Stokes Raman scattering (CARS) signal from liquid benzene was amplified with the help of surface plasmon polaritons supported on a flat Ag film antenna~\cite{shen79}. Subsequent progress on  surface-enhanced CRS has focused largely on engineering metallic nanoantennas based on Au and Ag to increase and/or optimize the sensitivity of detection~\cite{kumar2020time,fabelinsky2018surface,hayazawa2004amplification,ichimura2004tip,Liang1994,Voronine2012,steuwe2011surface,addison2009tuning,zhang2017high,fast2016surface,kenison2017imaging,zong2022wide,zong2021plasmon,frontiera2011surface,schlucker2011immuno}. The quest has seen considerable success and single-molecule sensitivity has been demonstrated by multiple groups~\cite{koo2005single,yampolsky2014seeing,zhang2014coherent,zong2019plasmon}. However, significant local heating at metallic junctions has proved to be a major obstacle, affecting the reproducibility of the SE-CARS measurements. When subjecting metallic nanoantennas to ultrafast irradiation, the local peak intensities can reach $\sim10^{12}-10^{13}$ W/cm$^2$, a range in which photo-induced ionization of the molecule can happen due to the bending of the potential followed by electron tunneling~\cite{keldysh1965ionization, sun1993femtosecond, groeneveld1995femtosecond, elsayed1991femtosecond, bouhelier2005surface, li2013landau, brongersma2015plasmon, alabastri2015high,dey2016observation,crampton2016ultrafast}. The morphological changes to the antenna due to intense local heating and plasmon-induced chemistry make it notoriously difficult to perform SE-CARS measurement in a reliable and reproducible manner~\cite{fast2019coherent}.  

The reproducibility challenges associated with SE-CARS measurements based on plasmonic nanoantennas originate from photo-induced heating effects that are intrinsic to metals. To make SE-CARS systems more reliable, moving beyond these lossy and fragile metallic substrates is necessary. High-index dielectric particles such as Si, Ge and GaP are known to exhibit favorable electromagnetic and thermal properties at the optical and near-infrared frequencies~\cite{caldarola2015non,albella2014electric,albella2013low,alessandri2016enhanced,barreda2019recent}. These properties make them particularly suitable for surface-enhanced spectroscopy applications, where heating of the antenna is a limiting factor. In nanoantennas based on these materials, localization of the incident optical fields occur as strong displacement currents are generated inside the material. These currents produce electromagnetic fields both inside and outside the material, which lead to confinement of the incident electromagnetic fields when nanocavities are formed between these dielectric particles. The potential of dielectric nano-structures to efficiently confine the electromagentic field has been known for a while and has been implemented in some SERS studies~\cite{alessandri2016enhanced,bontempi2014probing,wang2011using,khorasaninejad2012highly,khorasaninejad2012enhanced,huang2011enhanced}. We have recently shown that the use of dielectric materials in antenna designs can help mitigate photo-induced instabilities in SE-CARS measurements\cite{abedin2022}. That work focused on antennas formed between a Si nanoparticle and a flat Au film, which afforded reproducible SE-CARS signals, but suffered from a strong electronic two-photon excited luminescence (TPEL) background originating from the gold film.

In our current work, we present the first demonstration of SE-CARS measurements without the involvement of surface plasmons supported by metallic nanostructures. Instead, we use a Si-based all-dielectric nano-antenna with favorable thermal properties for stabilizing the CARS signal as well as to substantially reduce the TPEL background. Using all-dielectric nanojunctions, we show that the stability of the CARS signal is dramatically improved relative to the SE-CARS signal observed from Au nano-dumbbell based antennas. This work paves the way for future surface-enhanced coherent Raman sensors that offer the reproducibility needed for reliable high-speed biomolecular screening applications.

\section{Methods}

The silicon-based nano-antennas in this study are assembled from 150~nm diameter, pyridine-functionalized Si nanoparticles  (Meliorum Technologies), which are used without further purification. The nano-particles are suspended in water, sonicated and spin-cast on a BK7 coverslip prior to measurement. Nanojunctions are formed by spontaneous congregation of the particles during the spin-casting process. We use gold-based nano-antennas as a reference system for comparison. This antenna system has previously been used for SE-CARS experiments.\cite{yampolsky2014seeing,crampton2016ultrafast} The system consist of dimers of $\sim90$~nm diameter Au nanospheres, covered with trans 1,2-bis(4- pyridyl) ethylene (BPE), obtained from Cabot Security Materials. The nano-dumbbells are encapsulated in a silica shell of thickness 40-70~nm. The nano-dumbbells are suspended in water, sonicated and spin-cast onto BK7 glass coverslips. The SE-CARS spectrum of the BPE-functionalized Au dumbbell antenna shows two clear peaks in the 1560-1650~$\rm{cm}^{-1}$ range, attributed to the pyridyl ring stretching mode and the ethylene C=C stretching vibration (see Supporting Information).

\begin{figure}[htp]
    \centering
\includegraphics[width =3.25in]{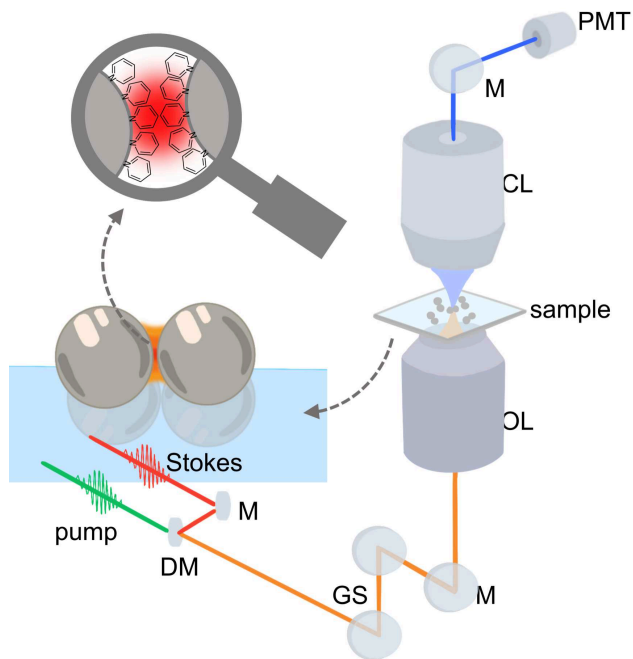}
    \caption{Schematic of the SE-CARS microscope. GS, galvanometric scanner; M, mirror; DM, dichroic mirror; OL, objective lens; CL, condenser lens; PMT, photomultiplier tube}
    \label{fig:f1}
\end{figure}

A schematic of the SE-CARS microscope system is shown in Figure \ref{fig:f1}. The SE-CARS system is based on a laser-scanning optical microscope (Fluoview 300, Olympus) with an Olympus IX71 inverted microscope frame. A modelocked fiber laser (aeroPULSE, NKT Photonics) with $\sim$2~ps long pulses and 80~MHz repetition rate provides the wavelength-fixed Stokes beam (1031~nm) for the SE-CARS experiments. A portion of this beam is frequency-doubled and the generated 515.5 nm beam synchronously pumps an optical parametric oscillator (Picoemerald, APE) to generate the tunable (700-960 nm) pump beam. The beams are spatially and temporally overlapped and subsequently conditioned to produce linearly polarized beams with clean transverse Gaussian profiles. The collinear beams are then focused using a water immersion 1.15 NA 40x objective lense (Olympus). The generated anti-Stokes signal is collimated using a condenser, filtered (775 $\pm$ 25 nm, Semrock) and detected using a photo-multiplier tube (R3896, Hamamatsu). In the experiments involving Si-particles, the average power for each beam is varied between 500 $\mu$W - 2 mW at the sample plane. A galvanometric scanner allows raster scanning over an 117 $\mu$m $\times$ 117 $\mu$m area of the sample, with a pixel dwell time of 4.27 $\mu$s per pixel. This corresponds to 1.12 frames/second acquisition rate for a $512\times512$ frame. For SE-CARS measurements on Au nano-dumbbells, the power for each of the incident beams is varied between 10 $\mu$W - 500 $\mu$W at the sample plane, while using the same dwell time of 4.27 $\mu$s per pixel.

Lumerical FDTD is used to perform local electromagnetic field calculations using the finite-difference time-domain (FDTD) method. Data for the dielectric function of all the materials are taken from the Palik library~\cite{edward1985handbook}. In the simulations, we consider nanojunctions formed between two Si nanoparticles, forming a dimer as the exemplar nanoantenna system. Fine meshing of 0.1~nm is used near the Si-dimer junction. The same geometry is used to calculate the absorption spectrum of the antenna system. Electromagnetic field enhancement data for the Au nano-dumbbell system are taken from Crampton et al.~\cite{crampton2016ultrafast}.

COMSOL Multiphysics is used to calculate the thermal response of both nanoantenna systems due to exposure to electromagnetic radiation. Water is chosen as the surrounding media in order to replicate actual experimental conditions. The Palik library is used for the thermal simulations as well.

\section{Results}

\subsection{SE-CARS with Si nanoantenna}
\begin{figure}[htp]
    \centering
\includegraphics[width =3.25in]{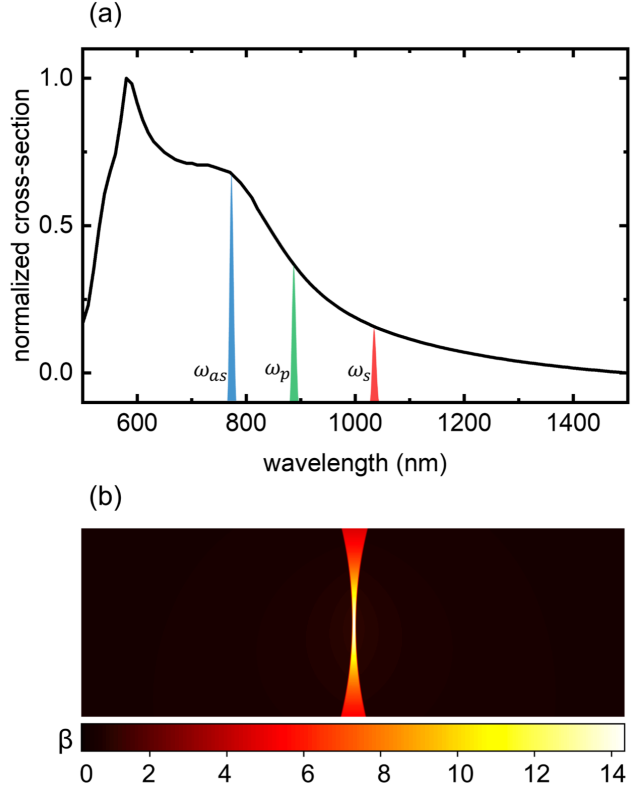}
    \caption{a) Calculated absorption spectrum of Si dimer nanoantenna system. The three colors represent the anti-Stokes, pump and Stokes wavelengths (from shorter to longer wavelengths). b) local electric field enhancement near the Si-Si nanojunction under 885 nm illumination. The local field enhancement factor is denoted by $\beta(\omega) = |E_{loc}/E_0$|}
    \label{fig:f2}
\end{figure}

In this work, we move away from metal-based nanoantenna systems and use dielectric antenna formed from clustered Si-nanoparticles to enhance the coherent Raman response of the molecule.  In order to examine the capability of this dielectric antenna to efficiently capture and confine incident optical fields, we use Lumerical FDTD to calculate the expected absorption spectrum of the antenna and the local fields inside the nanojunction. Figure \ref{fig:f2}a shows the calculated extinction spectrum of an antenna system from two Si-nanoparticles that form the exemplar dimer, along with the pump, Stokes and anti-Stokes frequencies involved in the experiment. The spectrum of the dielectric antenna is composed of a magnetic response in the optical range in addition to an electric resonance. This response is due to the displacement currents originating from the oscillation of bound electrons. The electric field calculations (shown in Figure \ref{fig:f2}b) reveal an effective field enhancement factor $\beta_p$ of $\sim$14 for the pump wavelength of 885~nm. Similar calculations are performed for the Stokes and anti-Stokes wavelengths resulting in $\beta_s = 13$ ($\lambda_{s}=1031~\rm{nm}$) and $\beta_{as}=15$ ($\lambda_{as}=775~\rm{nm}$). As such, the expected overall CARS enhancement factor is $\beta^4_{p}\beta^2_{s}\beta^2_{as} = 1.46\times 10^9$. The diameter of the electromagnetic hotspot is estimated as $\sim$6 nm. Assuming a surface coverage of 6~molecules/nm$^2$ and both particles have monolayers of pyridine, we are exciting $\sim$340 molecules in each junction.

\begin{figure}[htp]
    \centering
\includegraphics[width =3.25in]{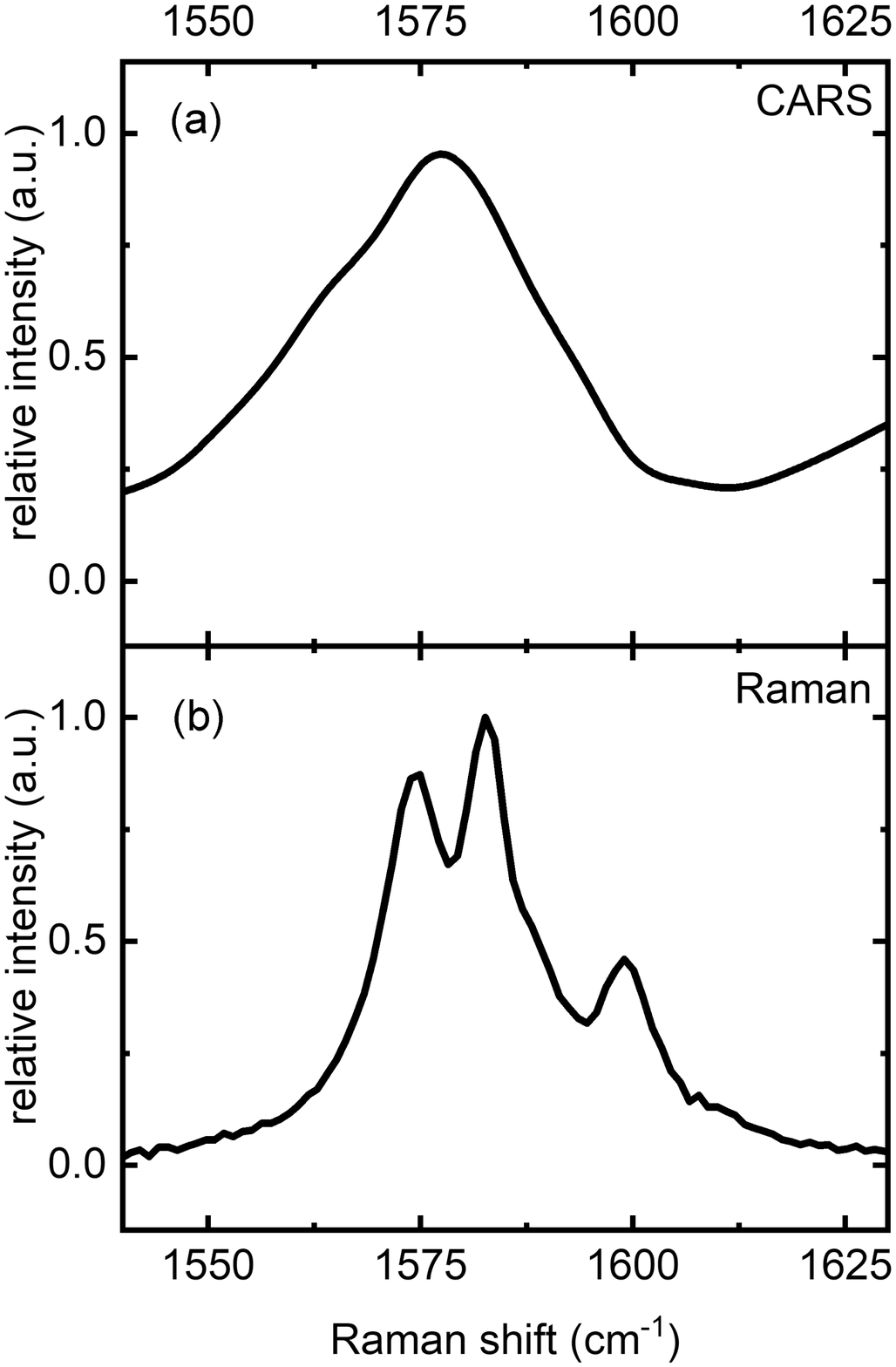}
    \caption{(a) CARS and (b) spontaneous Raman spectra of 99\% pyridine in the 1540-1630~$\rm{cm}^{-1}$ range.}
    \label{fig:f3}
\end{figure}

We target the ring stretching modes near 1580~cm$^{-1}$ for our SE-CARS experiment based on the CARS spectrum for 99\% pure pyridine, as seen in Figure \ref{fig:f3}(a). The spectral measurements are taken at 6 cm$^{-1}$ intervals, ranging from 1552 cm$^{-1}$ to 1630 cm$^{-1}$. This corresponds to scanning the pump wavelength between 882.7 nm and 888.8 nm. The spontaneous Raman spectrum of pyridine shown in Figure \ref{fig:f3}(b) reveals that the CARS signature derives from several Raman modes. The pyridine modes in this range are substantially weaker than the ring breathing modes near 1000~cm$^{-1}$~\cite{Benassi2021}, but the selected modes are targeted here because of better laser performance for the CARS process in this spectral range. Note that in addition to the resonant signature in the CARS spectrum, a nonresonant background from pyridine itself is present.

\begin{figure}[htp]
    \centering
\includegraphics[width =3.25in]{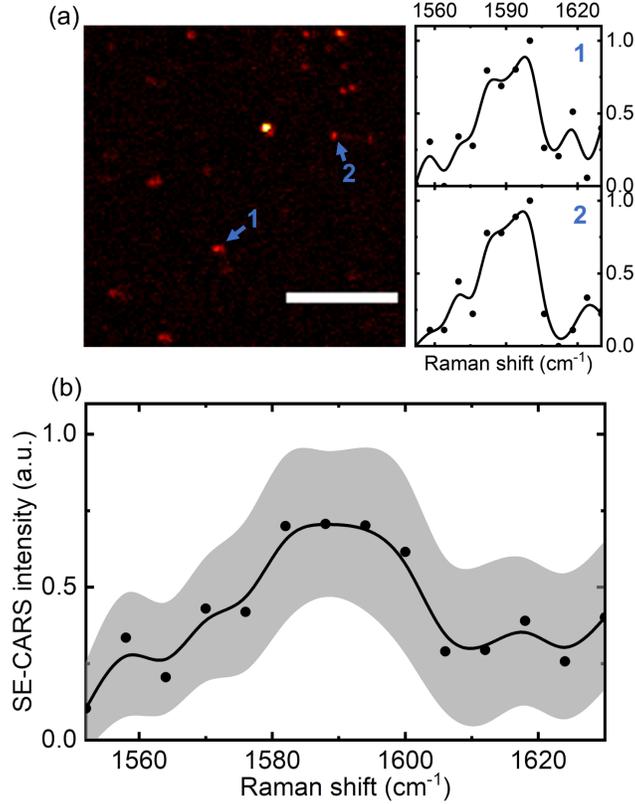}
    \caption{SE-CARS with Si nanoantenna. (a) SE-CARS image of pyridine functionalized Si nanoantenna system at 1582 cm$^{-1}$. Two representative spectra taken from locations 1 and 2. Scale bar is 10 $\mu$m. (b) Average of 50 SE-CARS spectra obtained from different locations. The line connecting the dots is used as a visual guide.}
    \label{fig:f4}
\end{figure}

In the dielectric nanoantenna system, consisting of clusters of Si particles, the measurements consistently produce vibrationally resonant signal from the target molecule. Background signal is acquired by temporally separating the pump and Stokes beams by~8 ps where no overlap is present. This background is subtracted to obtain the vibrationally resonant SE-CARS image. Figure \ref{fig:f4}a shows an SE-CARS image of clustered Si particles. The clusters appear as bright spots that are close to the diffraction-limited resolution of the SE-CARS microscope. Given the size of the spots, we estimate that these locations may each contain one to six 150~nm Si particles. Two sample locations (labeled 1 and 2 in Figure \ref{fig:f4}a) show the expected SE-CARS response from pyridine, suggesting the successful formation of a nanojunction, as evidenced by a distinct and reproducible peak in the 1580-1600 cm$^{-1}$ region. Figure \ref{fig:f4}b shows the mean SE-CARS spectrum with one standard deviation of error, obtained from 50 individual particle clusters. The spectrum shows clear vibrational sensitivity near the 1580-1600 cm$^{-1}$ spectral region. To the best of our knowledge, this is the first demonstration of surface-enhanced coherent Raman scattering with non-metallic antenna.

The CARS vibrational feature observed in the imaging experiments appears blue-shifted from the CARS signature seen in pure pyridine. Density functional theory calculations show that the pyridyl ring stretching vibration is expected to blue-shift upon bonding of the nitrogen to silicon (see Supporting Information). The blue-shift observed in the SE-CARS measurements thus supports the notion that the vibrational signature in the 1580-1600 cm$^{-1}$ range can be attributed to silicon-bonded pyridine molecules.

We note that there is a persistent background in most images taken in this configuration. The background signal is vibrationally non-resonant, depends on the spatial and temporal overlap of both beams and radiates at the ($2\omega_p - \omega_s$) frequency. Based on this observation, the origin of this background is determined to be four-wave mixing (FWM) due to the strong $\chi^{(3)}$ response of Si, which is found to contribute on average $\sim$24\% of the total signal on resonance. The contribution from broadband TPEL is on average $<$4\%.

\section{Thermal stability of dielectric and metal nanoantenna}

\begin{figure}[htp]
    \centering
\includegraphics[width =3.25in]{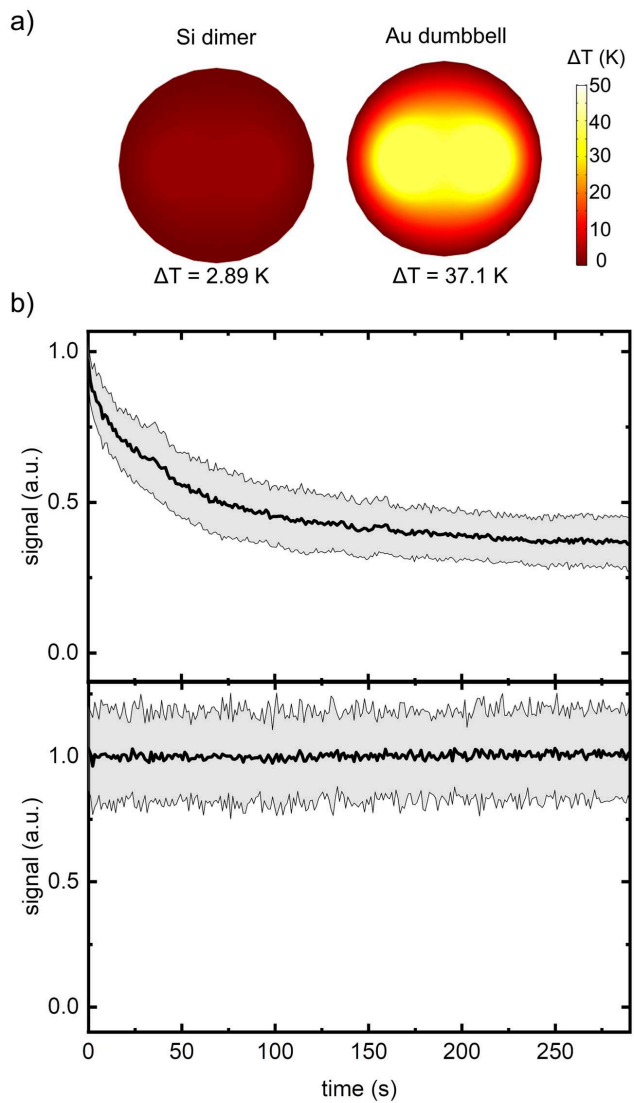}
    \caption{Thermal response of Si and Au nanoantenna systems. (a) FEM calculation of steady-state temperature of Si-dimer and Au nano-dumbbell antennas under 1 mW, 885 nm cw illumination. (b) Time evolution of experimental signal strength over 291 seconds for Au dumbbell (top) and Si-particle cluster (bottom) nanoantenna at their respective molecular resonance (1600 cm$^{-1}$ for Au dumbbell and 1588 cm$^{-1}$ for Si antenna system). The data represents averaged signal from 100 individual locations for each system.}
    \label{fig:f5}
\end{figure}

Figure \ref{fig:f5}a shows an FEM calculation (using COMSOL Multiphysics) of electromagnetic heating comparing the Si-dimer nanoantenna with a gold-based one. Both systems are immersed in water. The calculation demonstrates that under the exact same conditions and for the same geometrical dimension, heating of the Au antenna is $\sim$13 times higher (37.1 K) compared to heating of the Si (2.89 K) antenna. This simulated result is mirrored in experimental findings as well. We acquire SE-CARS images from ensembles of either Au or Si antenna systems for 291 seconds with 1.12 seconds per frame (4.27 $\mu$s/pixel). The incident power levels for both pump and Stokes beams were 2 mW at the sample plane for the Si antenna and 500 $\mu$W for Au antenna. While signal from the Au nano-dumbbell system decays rapidly and loses $\sim$60\% signal within 120 seconds (Figure \ref{fig:f5},top), the signal from the Si nanoantenna system remains robust with no noticeable decay in signal even after the end of the time series (Figure \ref{fig:f5},bottom). The numerical calculation and experimental observation of signal strength over time strongly supports our hypothesis that Si-based nonplasmonic nanoantennas can provide a new way to perform surface-enhanced coherent Raman scattering measurements with improved stability, reliability and reproducibility.

\section{Discussion}

In this work, we present the first experimental demonstration of surface-enhanced coherent Raman scattering of molecules with dielectric nanoantennas. Our recent work~\cite{abedin2022}, where we introduced metal-dielectric heterojunctions to improve thermal stability of the SE-CARS signal, motivated us to explore the use of non-metallic nanoantennas for SE-CARS measurements. The results discussed herein shows that it is possible to move beyond lossy and thermally unstable metallic systems and use all-dielectric nanoantennas for surface-enhanced coherent Raman scattering experiments instead. 

We have used spatially and temporally overlapped pump and Stokes beams in a laser-scanning microscope for the SE-CARS experiments. We successfully obtain stable and reproducible coherent Raman signal from our target molecule. In most of our measurements, we observe a persistent FWM background accompanying the molecular signal, due to the strong $\chi^{(3)}$ response of Si. The non-luminescent nature of the dielectric Si nanoparticles results in a much lower luminescent background compared to metal-based SE-CARS systems. The fields inside the dielectric dimer junctions are rather modest ($\beta \sim 14$) compared to what can be achieved using metallic nanoantennas. However, the overall CARS enhancement is substantial ($\sim10^9$) due the $\sim\beta^8$ dependence of the SE-CARS signal. Unlike previous SE-CARS antenna systems where morphological distortion of the structure due to photo-induced heating~\cite{crampton2016ultrafast,yampolsky2014seeing} or intense TPEL background ~\cite{abedin2022} limits the allowable illumination dosage, this work enables us to apply higher field intensities to enhance the molecular signal even further. The increased incident intensity can compensate for the lower values of $\beta$, generating comparable signals as previously observed with metal-based antenna systems.

Compared to SERS measurements, which scale as $\beta^4$, the SE-CARS approach benefits from a much larger overall enhancement effect, which makes dielectric antenna systems especially attractive for nonlinear Raman measurements. In this context, we note that we have performed regular SERS micro-spectroscopy measurements on the same Si-particle clusters. Using 532~nm or 785~nm excitation light of up to 10~mW, focused on a single cluster of pyridine-coated Si-particles, we have observed no vibrational SERS signatures of pyridine with integration times up to 10 s. We surmise, therefore, that the use of dielectric antennas may have a larger impact on the SE-CARS field compared to their impact on SERS applications.

We have chosen Si as our antenna material because of its high refractive index in the visible and near-infrared frequencies. As a result, Si based nanoantennas can efficiently capture and confine incident optical fields to nanoscopic regions. Furthermore, Si suffers less from the ohmic losses and adverse thermal effects that are intrinsic to metallic nanoantennas. This favorable thermal response is due to the higher specific heat of Si (5x higher compared to Au~\cite{perry1997perry}). The improved thermal stability is evidenced in experiments as well. In metallic systems, blinking-type behavior of the signal and the decay of signal over a short period of time is common. However, when we use dielectric antennas, such behavior is not observed.

While this work provides a solution to a longstanding limitation of traditional surface-enhanced CRS sensors, there are steps that can be taken to improve it even further.  Fabricating dielectric nanostructures with tailored electric and magnetic properties can make the nanoantenna more efficient and improve the reliability and repeatability of the signal. In addition, immobilized periodic enhancement structures can be employed for high throughput biomolecular screening purposes, an application where SE-CARS can provide much faster signal acquisition than possible with SERS based screening systems. Finally, red-shifting the pump and Stokes beams can reduce optical absorption to even lower levels. The lower absorption can help with the thermal response of the material, as well as reduce the electronic background.

\section{Conclusion}

We have presented an experimental demonstration of surface-enhanced coherent anti-Stokes Raman scattering (SE-CARS) of molecules with a dielectric Si nanoantenna. We have found that the thermal stability of the SE-CARS signal from such dielectric antennas is significantly better compared to corresponding signals from Au nano-dumbell antennas. This work shows that the inclusion of a dielectric nanoantenna can overcome the longstanding limitation of traditional metallic SE-CARS systems where unfavorable thermal properties adversely affect the stability and reproducibility of the measurements.

\bibliography{si_secars}

\begin{thebibliography}{55}%
\makeatletter
\providecommand \@ifxundefined [1]{%
 \@ifx{#1\undefined}
}%
\providecommand \@ifnum [1]{%
 \ifnum #1\expandafter \@firstoftwo
 \else \expandafter \@secondoftwo
 \fi
}%
\providecommand \@ifx [1]{%
 \ifx #1\expandafter \@firstoftwo
 \else \expandafter \@secondoftwo
 \fi
}%
\providecommand \natexlab [1]{#1}%
\providecommand \enquote  [1]{``#1''}%
\providecommand \bibnamefont  [1]{#1}%
\providecommand \bibfnamefont [1]{#1}%
\providecommand \citenamefont [1]{#1}%
\providecommand \href@noop [0]{\@secondoftwo}%
\providecommand \href [0]{\begingroup \@sanitize@url \@href}%
\providecommand \@href[1]{\@@startlink{#1}\@@href}%
\providecommand \@@href[1]{\endgroup#1\@@endlink}%
\providecommand \@sanitize@url [0]{\catcode `\\12\catcode `\$12\catcode
  `\&12\catcode `\#12\catcode `\^12\catcode `\_12\catcode `\%12\relax}%
\providecommand \@@startlink[1]{}%
\providecommand \@@endlink[0]{}%
\providecommand \url  [0]{\begingroup\@sanitize@url \@url }%
\providecommand \@url [1]{\endgroup\@href {#1}{\urlprefix }}%
\providecommand \urlprefix  [0]{URL }%
\providecommand \Eprint [0]{\href }%
\providecommand \doibase [0]{http://dx.doi.org/}%
\providecommand \selectlanguage [0]{\@gobble}%
\providecommand \bibinfo  [0]{\@secondoftwo}%
\providecommand \bibfield  [0]{\@secondoftwo}%
\providecommand \translation [1]{[#1]}%
\providecommand \BibitemOpen [0]{}%
\providecommand \bibitemStop [0]{}%
\providecommand \bibitemNoStop [0]{.\EOS\space}%
\providecommand \EOS [0]{\spacefactor3000\relax}%
\providecommand \BibitemShut  [1]{\csname bibitem#1\endcsname}%
\let\auto@bib@innerbib\@empty
\bibitem [{\citenamefont {McCreery}(2005)}]{mccreery2005raman}%
  \BibitemOpen
  \bibfield  {author} {\bibinfo {author} {\bibfnamefont {R.~L.}\ \bibnamefont
  {McCreery}},\ }\href@noop {} {\emph {\bibinfo {title} {Raman spectroscopy for
  chemical analysis}}},\ Vol.\ \bibinfo {volume} {225}\ (\bibinfo  {publisher}
  {John Wiley \& Sons},\ \bibinfo {year} {2005})\BibitemShut {NoStop}%
\bibitem [{\citenamefont {Fast}\ and\ \citenamefont
  {Potma}(2019)}]{fast2019coherent}%
  \BibitemOpen
  \bibfield  {author} {\bibinfo {author} {\bibfnamefont {A.}~\bibnamefont
  {Fast}}\ and\ \bibinfo {author} {\bibfnamefont {E.~O.}\ \bibnamefont
  {Potma}},\ }\href@noop {} {\bibfield  {journal} {\bibinfo  {journal}
  {Nanophotonics}\ }\textbf {\bibinfo {volume} {8}},\ \bibinfo {pages} {991}
  (\bibinfo {year} {2019})}\BibitemShut {NoStop}%
\bibitem [{\citenamefont {Moskovits}(1985)}]{Moskovits1985}%
  \BibitemOpen
  \bibfield  {author} {\bibinfo {author} {\bibfnamefont {M.}~\bibnamefont
  {Moskovits}},\ }\href {https://link.aps.org/doi/10.1103/RevModPhys.57.783}
  {\bibfield  {journal} {\bibinfo  {journal} {Rev. Mod. Phys.}\ }\textbf
  {\bibinfo {volume} {57}},\ \bibinfo {pages} {783} (\bibinfo {year}
  {1985})}\BibitemShut {NoStop}%
\bibitem [{\citenamefont {Willets}\ and\ \citenamefont
  {Van~Duyne}(2007)}]{willets2007localized}%
  \BibitemOpen
  \bibfield  {author} {\bibinfo {author} {\bibfnamefont {K.~A.}\ \bibnamefont
  {Willets}}\ and\ \bibinfo {author} {\bibfnamefont {R.~P.}\ \bibnamefont
  {Van~Duyne}},\ }\href@noop {} {\bibfield  {journal} {\bibinfo  {journal}
  {Annu. Rev. Phys. Chem.}\ }\textbf {\bibinfo {volume} {58}},\ \bibinfo
  {pages} {267} (\bibinfo {year} {2007})}\BibitemShut {NoStop}%
\bibitem [{\citenamefont {Zhang}\ \emph {et~al.}(2013)\citenamefont {Zhang},
  \citenamefont {Zhang}, \citenamefont {Dong}, \citenamefont {Jiang},
  \citenamefont {Zhang}, \citenamefont {Chen}, \citenamefont {Zhang},
  \citenamefont {Liao}, \citenamefont {Aizpurua}, \citenamefont {Luo} \emph
  {et~al.}}]{zhang2013chemical}%
  \BibitemOpen
  \bibfield  {author} {\bibinfo {author} {\bibfnamefont {R.}~\bibnamefont
  {Zhang}}, \bibinfo {author} {\bibfnamefont {Y.}~\bibnamefont {Zhang}},
  \bibinfo {author} {\bibfnamefont {Z.}~\bibnamefont {Dong}}, \bibinfo {author}
  {\bibfnamefont {S.}~\bibnamefont {Jiang}}, \bibinfo {author} {\bibfnamefont
  {C.}~\bibnamefont {Zhang}}, \bibinfo {author} {\bibfnamefont
  {L.}~\bibnamefont {Chen}}, \bibinfo {author} {\bibfnamefont {L.}~\bibnamefont
  {Zhang}}, \bibinfo {author} {\bibfnamefont {Y.}~\bibnamefont {Liao}},
  \bibinfo {author} {\bibfnamefont {J.}~\bibnamefont {Aizpurua}}, \bibinfo
  {author} {\bibfnamefont {Y.~e.}\ \bibnamefont {Luo}},  \emph {et~al.},\
  }\href@noop {} {\bibfield  {journal} {\bibinfo  {journal} {Nature}\ }\textbf
  {\bibinfo {volume} {498}},\ \bibinfo {pages} {82} (\bibinfo {year}
  {2013})}\BibitemShut {NoStop}%
\bibitem [{\citenamefont {Fleischmann}\ \emph {et~al.}(1974)\citenamefont
  {Fleischmann}, \citenamefont {Hendra},\ and\ \citenamefont
  {McQuillan}}]{fleischmann1974raman}%
  \BibitemOpen
  \bibfield  {author} {\bibinfo {author} {\bibfnamefont {M.}~\bibnamefont
  {Fleischmann}}, \bibinfo {author} {\bibfnamefont {P.~J.}\ \bibnamefont
  {Hendra}}, \ and\ \bibinfo {author} {\bibfnamefont {A.~J.}\ \bibnamefont
  {McQuillan}},\ }\href@noop {} {\bibfield  {journal} {\bibinfo  {journal}
  {Chemical physics letters}\ }\textbf {\bibinfo {volume} {26}},\ \bibinfo
  {pages} {163} (\bibinfo {year} {1974})}\BibitemShut {NoStop}%
\bibitem [{\citenamefont {Nie}\ and\ \citenamefont
  {Emory}(1997)}]{nie1997probing}%
  \BibitemOpen
  \bibfield  {author} {\bibinfo {author} {\bibfnamefont {S.}~\bibnamefont
  {Nie}}\ and\ \bibinfo {author} {\bibfnamefont {S.~R.}\ \bibnamefont
  {Emory}},\ }\href@noop {} {\bibfield  {journal} {\bibinfo  {journal}
  {science}\ }\textbf {\bibinfo {volume} {275}},\ \bibinfo {pages} {1102}
  (\bibinfo {year} {1997})}\BibitemShut {NoStop}%
\bibitem [{\citenamefont {Le~Ru}\ and\ \citenamefont
  {Etchegoin}(2012)}]{le2012single}%
  \BibitemOpen
  \bibfield  {author} {\bibinfo {author} {\bibfnamefont {E.~C.}\ \bibnamefont
  {Le~Ru}}\ and\ \bibinfo {author} {\bibfnamefont {P.~G.}\ \bibnamefont
  {Etchegoin}},\ }\href@noop {} {\bibfield  {journal} {\bibinfo  {journal}
  {Annual review of physical chemistry}\ }\textbf {\bibinfo {volume} {63}},\
  \bibinfo {pages} {65} (\bibinfo {year} {2012})}\BibitemShut {NoStop}%
\bibitem [{\citenamefont {Dieringer}\ \emph {et~al.}(2007)\citenamefont
  {Dieringer}, \citenamefont {Lettan}, \citenamefont {Scheidt},\ and\
  \citenamefont {Van~Duyne}}]{Dieringer2007}%
  \BibitemOpen
  \bibfield  {author} {\bibinfo {author} {\bibfnamefont {J.~A.}\ \bibnamefont
  {Dieringer}}, \bibinfo {author} {\bibfnamefont {R.~B.}\ \bibnamefont
  {Lettan}}, \bibinfo {author} {\bibfnamefont {K.~A.}\ \bibnamefont {Scheidt}},
  \ and\ \bibinfo {author} {\bibfnamefont {R.~P.}\ \bibnamefont {Van~Duyne}},\
  }\href {\doibase 10.1021/ja077243c} {\bibfield  {journal} {\bibinfo
  {journal} {Journal of the American Chemical Society}\ }\textbf {\bibinfo
  {volume} {129}},\ \bibinfo {pages} {16249} (\bibinfo {year}
  {2007})}\BibitemShut {NoStop}%
\bibitem [{\citenamefont {Kneipp}\ \emph {et~al.}(1998)\citenamefont {Kneipp},
  \citenamefont {Kneipp}, \citenamefont {Kartha}, \citenamefont {Manoharan},
  \citenamefont {Deinum}, \citenamefont {Itzkan}, \citenamefont {Dasari},\ and\
  \citenamefont {Feld}}]{Kneipp1998_nucleo}%
  \BibitemOpen
  \bibfield  {author} {\bibinfo {author} {\bibfnamefont {K.}~\bibnamefont
  {Kneipp}}, \bibinfo {author} {\bibfnamefont {H.}~\bibnamefont {Kneipp}},
  \bibinfo {author} {\bibfnamefont {V.~B.}\ \bibnamefont {Kartha}}, \bibinfo
  {author} {\bibfnamefont {R.}~\bibnamefont {Manoharan}}, \bibinfo {author}
  {\bibfnamefont {G.}~\bibnamefont {Deinum}}, \bibinfo {author} {\bibfnamefont
  {I.}~\bibnamefont {Itzkan}}, \bibinfo {author} {\bibfnamefont {R.~R.}\
  \bibnamefont {Dasari}}, \ and\ \bibinfo {author} {\bibfnamefont {M.~S.}\
  \bibnamefont {Feld}},\ }\href {\doibase 10.1103/PhysRevE.57.R6281} {\bibfield
   {journal} {\bibinfo  {journal} {Phys. Rev. E}\ }\textbf {\bibinfo {volume}
  {57}},\ \bibinfo {pages} {R6281} (\bibinfo {year} {1998})}\BibitemShut
  {NoStop}%
\bibitem [{\citenamefont {Chen}\ \emph {et~al.}(2018)\citenamefont {Chen},
  \citenamefont {Li}, \citenamefont {Kerman}, \citenamefont {Neutens},
  \citenamefont {Willems}, \citenamefont {Cornelissen}, \citenamefont {Lagae},
  \citenamefont {Stakenborg},\ and\ \citenamefont {Dorpe}}]{Chen2018}%
  \BibitemOpen
  \bibfield  {author} {\bibinfo {author} {\bibfnamefont {C.}~\bibnamefont
  {Chen}}, \bibinfo {author} {\bibfnamefont {Y.}~\bibnamefont {Li}}, \bibinfo
  {author} {\bibfnamefont {S.}~\bibnamefont {Kerman}}, \bibinfo {author}
  {\bibfnamefont {P.}~\bibnamefont {Neutens}}, \bibinfo {author} {\bibfnamefont
  {K.}~\bibnamefont {Willems}}, \bibinfo {author} {\bibfnamefont
  {S.}~\bibnamefont {Cornelissen}}, \bibinfo {author} {\bibfnamefont
  {L.}~\bibnamefont {Lagae}}, \bibinfo {author} {\bibfnamefont
  {T.}~\bibnamefont {Stakenborg}}, \ and\ \bibinfo {author} {\bibfnamefont
  {P.~V.}\ \bibnamefont {Dorpe}},\ }\href@noop {} {\bibfield  {journal}
  {\bibinfo  {journal} {Nature Communications}\ }\textbf {\bibinfo {volume}
  {9}},\ \bibinfo {pages} {1733} (\bibinfo {year} {2018})}\BibitemShut
  {NoStop}%
\bibitem [{\citenamefont {Chen}\ \emph {et~al.}(1979)\citenamefont {Chen},
  \citenamefont {de~Castro}, \citenamefont {Shen},\ and\ \citenamefont
  {DeMartini}}]{shen79}%
  \BibitemOpen
  \bibfield  {author} {\bibinfo {author} {\bibfnamefont {C.~K.}\ \bibnamefont
  {Chen}}, \bibinfo {author} {\bibfnamefont {A.~R.~B.}\ \bibnamefont
  {de~Castro}}, \bibinfo {author} {\bibfnamefont {Y.~R.}\ \bibnamefont {Shen}},
  \ and\ \bibinfo {author} {\bibfnamefont {F.}~\bibnamefont {DeMartini}},\
  }\href {\doibase 10.1103/PhysRevLett.43.946} {\bibfield  {journal} {\bibinfo
  {journal} {Phys. Rev. Lett.}\ }\textbf {\bibinfo {volume} {43}},\ \bibinfo
  {pages} {946} (\bibinfo {year} {1979})}\BibitemShut {NoStop}%
\bibitem [{\citenamefont {Kumar}\ \emph {et~al.}(2020)\citenamefont {Kumar},
  \citenamefont {Kuramochi}, \citenamefont {Takeuchi},\ and\ \citenamefont
  {Tahara}}]{kumar2020time}%
  \BibitemOpen
  \bibfield  {author} {\bibinfo {author} {\bibfnamefont {P.}~\bibnamefont
  {Kumar}}, \bibinfo {author} {\bibfnamefont {H.}~\bibnamefont {Kuramochi}},
  \bibinfo {author} {\bibfnamefont {S.}~\bibnamefont {Takeuchi}}, \ and\
  \bibinfo {author} {\bibfnamefont {T.}~\bibnamefont {Tahara}},\ }\href@noop {}
  {\bibfield  {journal} {\bibinfo  {journal} {The Journal of Physical Chemistry
  Letters}\ }\textbf {\bibinfo {volume} {11}},\ \bibinfo {pages} {6305}
  (\bibinfo {year} {2020})}\BibitemShut {NoStop}%
\bibitem [{\citenamefont {Fabelinsky}\ \emph {et~al.}(2018)\citenamefont
  {Fabelinsky}, \citenamefont {Kozlov}, \citenamefont {Orlov}, \citenamefont
  {Polivanov}, \citenamefont {Shcherbakov}, \citenamefont {Smirnov},
  \citenamefont {Vereschagin}, \citenamefont {Arzumanyan}, \citenamefont
  {Mamatkulov}, \citenamefont {Afanasiev} \emph
  {et~al.}}]{fabelinsky2018surface}%
  \BibitemOpen
  \bibfield  {author} {\bibinfo {author} {\bibfnamefont {V.~I.}\ \bibnamefont
  {Fabelinsky}}, \bibinfo {author} {\bibfnamefont {D.~N.}\ \bibnamefont
  {Kozlov}}, \bibinfo {author} {\bibfnamefont {S.~N.}\ \bibnamefont {Orlov}},
  \bibinfo {author} {\bibfnamefont {Y.~N.}\ \bibnamefont {Polivanov}}, \bibinfo
  {author} {\bibfnamefont {I.~A.}\ \bibnamefont {Shcherbakov}}, \bibinfo
  {author} {\bibfnamefont {V.~V.}\ \bibnamefont {Smirnov}}, \bibinfo {author}
  {\bibfnamefont {K.~A.}\ \bibnamefont {Vereschagin}}, \bibinfo {author}
  {\bibfnamefont {G.~M.}\ \bibnamefont {Arzumanyan}}, \bibinfo {author}
  {\bibfnamefont {K.~Z.}\ \bibnamefont {Mamatkulov}}, \bibinfo {author}
  {\bibfnamefont {K.~N.}\ \bibnamefont {Afanasiev}},  \emph {et~al.},\
  }\href@noop {} {\bibfield  {journal} {\bibinfo  {journal} {Journal of Raman
  Spectroscopy}\ }\textbf {\bibinfo {volume} {49}},\ \bibinfo {pages} {1145}
  (\bibinfo {year} {2018})}\BibitemShut {NoStop}%
\bibitem [{\citenamefont {Hayazawa}\ \emph {et~al.}(2004)\citenamefont
  {Hayazawa}, \citenamefont {Ichimura}, \citenamefont {Hashimoto},
  \citenamefont {Inouye},\ and\ \citenamefont
  {Kawata}}]{hayazawa2004amplification}%
  \BibitemOpen
  \bibfield  {author} {\bibinfo {author} {\bibfnamefont {N.}~\bibnamefont
  {Hayazawa}}, \bibinfo {author} {\bibfnamefont {T.}~\bibnamefont {Ichimura}},
  \bibinfo {author} {\bibfnamefont {M.}~\bibnamefont {Hashimoto}}, \bibinfo
  {author} {\bibfnamefont {Y.}~\bibnamefont {Inouye}}, \ and\ \bibinfo {author}
  {\bibfnamefont {S.}~\bibnamefont {Kawata}},\ }\href@noop {} {\bibfield
  {journal} {\bibinfo  {journal} {Journal of applied physics}\ }\textbf
  {\bibinfo {volume} {95}},\ \bibinfo {pages} {2676} (\bibinfo {year}
  {2004})}\BibitemShut {NoStop}%
\bibitem [{\citenamefont {Ichimura}\ \emph {et~al.}(2004)\citenamefont
  {Ichimura}, \citenamefont {Hayazawa}, \citenamefont {Hashimoto},
  \citenamefont {Inouye},\ and\ \citenamefont {Kawata}}]{ichimura2004tip}%
  \BibitemOpen
  \bibfield  {author} {\bibinfo {author} {\bibfnamefont {T.}~\bibnamefont
  {Ichimura}}, \bibinfo {author} {\bibfnamefont {N.}~\bibnamefont {Hayazawa}},
  \bibinfo {author} {\bibfnamefont {M.}~\bibnamefont {Hashimoto}}, \bibinfo
  {author} {\bibfnamefont {Y.}~\bibnamefont {Inouye}}, \ and\ \bibinfo {author}
  {\bibfnamefont {S.}~\bibnamefont {Kawata}},\ }\href@noop {} {\bibfield
  {journal} {\bibinfo  {journal} {Physical review letters}\ }\textbf {\bibinfo
  {volume} {92}},\ \bibinfo {pages} {220801} (\bibinfo {year}
  {2004})}\BibitemShut {NoStop}%
\bibitem [{\citenamefont {Liang}\ \emph {et~al.}(1994)\citenamefont {Liang},
  \citenamefont {Weippert}, \citenamefont {Funk}, \citenamefont {Materny},\
  and\ \citenamefont {Kiefer}}]{Liang1994}%
  \BibitemOpen
  \bibfield  {author} {\bibinfo {author} {\bibfnamefont {E.}~\bibnamefont
  {Liang}}, \bibinfo {author} {\bibfnamefont {A.}~\bibnamefont {Weippert}},
  \bibinfo {author} {\bibfnamefont {J.-M.}\ \bibnamefont {Funk}}, \bibinfo
  {author} {\bibfnamefont {A.}~\bibnamefont {Materny}}, \ and\ \bibinfo
  {author} {\bibfnamefont {W.}~\bibnamefont {Kiefer}},\ }\href {\doibase
  https://doi.org/10.1016/0009-2614(94)00779-9} {\bibfield  {journal} {\bibinfo
   {journal} {Chemical Physics Letters}\ }\textbf {\bibinfo {volume} {227}},\
  \bibinfo {pages} {115 } (\bibinfo {year} {1994})}\BibitemShut {NoStop}%
\bibitem [{\citenamefont {Voronine}\ \emph {et~al.}(2012)\citenamefont
  {Voronine}, \citenamefont {Sinyokov}, \citenamefont {Hua}, \citenamefont
  {Wang}, \citenamefont {Jha}, \citenamefont {Munusamy}, \citenamefont
  {Wheeler}, \citenamefont {Welch}, \citenamefont {Sokolov},\ and\
  \citenamefont {Scully}}]{Voronine2012}%
  \BibitemOpen
  \bibfield  {author} {\bibinfo {author} {\bibfnamefont {D.~V.}\ \bibnamefont
  {Voronine}}, \bibinfo {author} {\bibfnamefont {A.~M.}\ \bibnamefont
  {Sinyokov}}, \bibinfo {author} {\bibfnamefont {X.}~\bibnamefont {Hua}},
  \bibinfo {author} {\bibfnamefont {K.}~\bibnamefont {Wang}}, \bibinfo {author}
  {\bibfnamefont {P.~K.}\ \bibnamefont {Jha}}, \bibinfo {author} {\bibfnamefont
  {E.}~\bibnamefont {Munusamy}}, \bibinfo {author} {\bibfnamefont {S.~E.}\
  \bibnamefont {Wheeler}}, \bibinfo {author} {\bibfnamefont {G.}~\bibnamefont
  {Welch}}, \bibinfo {author} {\bibfnamefont {A.}~\bibnamefont {Sokolov}}, \
  and\ \bibinfo {author} {\bibfnamefont {M.~O.}\ \bibnamefont {Scully}},\
  }\href@noop {} {\bibfield  {journal} {\bibinfo  {journal} {Sci. Rep.}\
  }\textbf {\bibinfo {volume} {2}},\ \bibinfo {pages} {891} (\bibinfo {year}
  {2012})}\BibitemShut {NoStop}%
\bibitem [{\citenamefont {Steuwe}\ \emph {et~al.}(2011)\citenamefont {Steuwe},
  \citenamefont {Kaminski}, \citenamefont {Baumberg},\ and\ \citenamefont
  {Mahajan}}]{steuwe2011surface}%
  \BibitemOpen
  \bibfield  {author} {\bibinfo {author} {\bibfnamefont {C.}~\bibnamefont
  {Steuwe}}, \bibinfo {author} {\bibfnamefont {C.~F.}\ \bibnamefont
  {Kaminski}}, \bibinfo {author} {\bibfnamefont {J.~J.}\ \bibnamefont
  {Baumberg}}, \ and\ \bibinfo {author} {\bibfnamefont {S.}~\bibnamefont
  {Mahajan}},\ }\href@noop {} {\bibfield  {journal} {\bibinfo  {journal} {Nano
  letters}\ }\textbf {\bibinfo {volume} {11}},\ \bibinfo {pages} {5339}
  (\bibinfo {year} {2011})}\BibitemShut {NoStop}%
\bibitem [{\citenamefont {Addison}\ \emph {et~al.}(2009)\citenamefont
  {Addison}, \citenamefont {Konorov}, \citenamefont {Brolo}, \citenamefont
  {Blades},\ and\ \citenamefont {Turner}}]{addison2009tuning}%
  \BibitemOpen
  \bibfield  {author} {\bibinfo {author} {\bibfnamefont {C.~J.}\ \bibnamefont
  {Addison}}, \bibinfo {author} {\bibfnamefont {S.~O.}\ \bibnamefont
  {Konorov}}, \bibinfo {author} {\bibfnamefont {A.~G.}\ \bibnamefont {Brolo}},
  \bibinfo {author} {\bibfnamefont {M.~W.}\ \bibnamefont {Blades}}, \ and\
  \bibinfo {author} {\bibfnamefont {R.~F.}\ \bibnamefont {Turner}},\
  }\href@noop {} {\bibfield  {journal} {\bibinfo  {journal} {The Journal of
  Physical Chemistry C}\ }\textbf {\bibinfo {volume} {113}},\ \bibinfo {pages}
  {3586} (\bibinfo {year} {2009})}\BibitemShut {NoStop}%
\bibitem [{\citenamefont {Zhang}\ and\ \citenamefont
  {Song}(2017)}]{zhang2017high}%
  \BibitemOpen
  \bibfield  {author} {\bibinfo {author} {\bibfnamefont {Z.}~\bibnamefont
  {Zhang}}\ and\ \bibinfo {author} {\bibfnamefont {G.}~\bibnamefont {Song}},\
  }\href@noop {} {\bibfield  {journal} {\bibinfo  {journal} {Journal of
  Semiconductors}\ }\textbf {\bibinfo {volume} {38}},\ \bibinfo {pages}
  {022001} (\bibinfo {year} {2017})}\BibitemShut {NoStop}%
\bibitem [{\citenamefont {Fast}\ \emph {et~al.}(2016)\citenamefont {Fast},
  \citenamefont {Kenison}, \citenamefont {Syme},\ and\ \citenamefont
  {Potma}}]{fast2016surface}%
  \BibitemOpen
  \bibfield  {author} {\bibinfo {author} {\bibfnamefont {A.}~\bibnamefont
  {Fast}}, \bibinfo {author} {\bibfnamefont {J.~P.}\ \bibnamefont {Kenison}},
  \bibinfo {author} {\bibfnamefont {C.~D.}\ \bibnamefont {Syme}}, \ and\
  \bibinfo {author} {\bibfnamefont {E.~O.}\ \bibnamefont {Potma}},\ }\href@noop
  {} {\bibfield  {journal} {\bibinfo  {journal} {Applied optics}\ }\textbf
  {\bibinfo {volume} {55}},\ \bibinfo {pages} {5994} (\bibinfo {year}
  {2016})}\BibitemShut {NoStop}%
\bibitem [{\citenamefont {Kenison}\ \emph {et~al.}(2017)\citenamefont
  {Kenison}, \citenamefont {Fast}, \citenamefont {Guo}, \citenamefont {LeBon},
  \citenamefont {Jiang},\ and\ \citenamefont {Potma}}]{kenison2017imaging}%
  \BibitemOpen
  \bibfield  {author} {\bibinfo {author} {\bibfnamefont {J.~P.}\ \bibnamefont
  {Kenison}}, \bibinfo {author} {\bibfnamefont {A.}~\bibnamefont {Fast}},
  \bibinfo {author} {\bibfnamefont {F.}~\bibnamefont {Guo}}, \bibinfo {author}
  {\bibfnamefont {A.}~\bibnamefont {LeBon}}, \bibinfo {author} {\bibfnamefont
  {W.}~\bibnamefont {Jiang}}, \ and\ \bibinfo {author} {\bibfnamefont {E.~O.}\
  \bibnamefont {Potma}},\ }\href@noop {} {\bibfield  {journal} {\bibinfo
  {journal} {JOSA B}\ }\textbf {\bibinfo {volume} {34}},\ \bibinfo {pages}
  {2104} (\bibinfo {year} {2017})}\BibitemShut {NoStop}%
\bibitem [{\citenamefont {Zong}\ \emph {et~al.}(2022)\citenamefont {Zong},
  \citenamefont {Cheng}, \citenamefont {Chen}, \citenamefont {Lin},
  \citenamefont {Zhang}, \citenamefont {Chen}, \citenamefont {Li},
  \citenamefont {Yang},\ and\ \citenamefont {Cheng}}]{zong2022wide}%
  \BibitemOpen
  \bibfield  {author} {\bibinfo {author} {\bibfnamefont {C.}~\bibnamefont
  {Zong}}, \bibinfo {author} {\bibfnamefont {R.}~\bibnamefont {Cheng}},
  \bibinfo {author} {\bibfnamefont {F.}~\bibnamefont {Chen}}, \bibinfo {author}
  {\bibfnamefont {P.}~\bibnamefont {Lin}}, \bibinfo {author} {\bibfnamefont
  {M.}~\bibnamefont {Zhang}}, \bibinfo {author} {\bibfnamefont
  {Z.}~\bibnamefont {Chen}}, \bibinfo {author} {\bibfnamefont {C.}~\bibnamefont
  {Li}}, \bibinfo {author} {\bibfnamefont {C.}~\bibnamefont {Yang}}, \ and\
  \bibinfo {author} {\bibfnamefont {J.-X.}\ \bibnamefont {Cheng}},\ }\href@noop
  {} {\bibfield  {journal} {\bibinfo  {journal} {ACS Photonics}\ }\textbf
  {\bibinfo {volume} {9}},\ \bibinfo {pages} {1042} (\bibinfo {year}
  {2022})}\BibitemShut {NoStop}%
\bibitem [{\citenamefont {Zong}\ \emph {et~al.}(2021)\citenamefont {Zong},
  \citenamefont {Xie}, \citenamefont {Zhang}, \citenamefont {Huang},
  \citenamefont {Yang},\ and\ \citenamefont {Cheng}}]{zong2021plasmon}%
  \BibitemOpen
  \bibfield  {author} {\bibinfo {author} {\bibfnamefont {C.}~\bibnamefont
  {Zong}}, \bibinfo {author} {\bibfnamefont {Y.}~\bibnamefont {Xie}}, \bibinfo
  {author} {\bibfnamefont {M.}~\bibnamefont {Zhang}}, \bibinfo {author}
  {\bibfnamefont {Y.}~\bibnamefont {Huang}}, \bibinfo {author} {\bibfnamefont
  {C.}~\bibnamefont {Yang}}, \ and\ \bibinfo {author} {\bibfnamefont {J.-X.}\
  \bibnamefont {Cheng}},\ }\href@noop {} {\bibfield  {journal} {\bibinfo
  {journal} {The Journal of Chemical Physics}\ }\textbf {\bibinfo {volume}
  {154}},\ \bibinfo {pages} {034201} (\bibinfo {year} {2021})}\BibitemShut
  {NoStop}%
\bibitem [{\citenamefont {Frontiera}\ \emph {et~al.}(2011)\citenamefont
  {Frontiera}, \citenamefont {Henry}, \citenamefont {Gruenke},\ and\
  \citenamefont {Van~Duyne}}]{frontiera2011surface}%
  \BibitemOpen
  \bibfield  {author} {\bibinfo {author} {\bibfnamefont {R.~R.}\ \bibnamefont
  {Frontiera}}, \bibinfo {author} {\bibfnamefont {A.-I.}\ \bibnamefont
  {Henry}}, \bibinfo {author} {\bibfnamefont {N.~L.}\ \bibnamefont {Gruenke}},
  \ and\ \bibinfo {author} {\bibfnamefont {R.~P.}\ \bibnamefont {Van~Duyne}},\
  }\href@noop {} {\bibfield  {journal} {\bibinfo  {journal} {The journal of
  physical chemistry letters}\ }\textbf {\bibinfo {volume} {2}},\ \bibinfo
  {pages} {1199} (\bibinfo {year} {2011})}\BibitemShut {NoStop}%
\bibitem [{\citenamefont {Schl\"ucker}\ \emph {et~al.}(2011)\citenamefont
  {Schl\"ucker}, \citenamefont {Salehi}, \citenamefont {Bergner}, \citenamefont
  {Sch\"utz}, \citenamefont {Str\"obel}, \citenamefont {Marx}, \citenamefont
  {Petersen}, \citenamefont {Dietzek},\ and\ \citenamefont
  {Popp}}]{schlucker2011immuno}%
  \BibitemOpen
  \bibfield  {author} {\bibinfo {author} {\bibfnamefont {S.}~\bibnamefont
  {Schl\"ucker}}, \bibinfo {author} {\bibfnamefont {M.}~\bibnamefont {Salehi}},
  \bibinfo {author} {\bibfnamefont {G.}~\bibnamefont {Bergner}}, \bibinfo
  {author} {\bibfnamefont {M.}~\bibnamefont {Sch\"utz}}, \bibinfo {author}
  {\bibfnamefont {P.}~\bibnamefont {Str\"obel}}, \bibinfo {author}
  {\bibfnamefont {A.}~\bibnamefont {Marx}}, \bibinfo {author} {\bibfnamefont
  {I.}~\bibnamefont {Petersen}}, \bibinfo {author} {\bibfnamefont
  {B.}~\bibnamefont {Dietzek}}, \ and\ \bibinfo {author} {\bibfnamefont
  {J.}~\bibnamefont {Popp}},\ }\href {\doibase 10.1021/ac201284d} {\bibfield
  {journal} {\bibinfo  {journal} {Analytical Chemistry}\ }\textbf {\bibinfo
  {volume} {83}},\ \bibinfo {pages} {7081} (\bibinfo {year}
  {2011})}\BibitemShut {NoStop}%
\bibitem [{\citenamefont {Koo}\ \emph {et~al.}(2005)\citenamefont {Koo},
  \citenamefont {Chan},\ and\ \citenamefont {Berlin}}]{koo2005single}%
  \BibitemOpen
  \bibfield  {author} {\bibinfo {author} {\bibfnamefont {T.-W.}\ \bibnamefont
  {Koo}}, \bibinfo {author} {\bibfnamefont {S.}~\bibnamefont {Chan}}, \ and\
  \bibinfo {author} {\bibfnamefont {A.~A.}\ \bibnamefont {Berlin}},\
  }\href@noop {} {\bibfield  {journal} {\bibinfo  {journal} {Optics letters}\
  }\textbf {\bibinfo {volume} {30}},\ \bibinfo {pages} {1024} (\bibinfo {year}
  {2005})}\BibitemShut {NoStop}%
\bibitem [{\citenamefont {Yampolsky}\ \emph {et~al.}(2014)\citenamefont
  {Yampolsky}, \citenamefont {Fishman}, \citenamefont {Dey}, \citenamefont
  {Hulkko}, \citenamefont {Banik}, \citenamefont {Potma},\ and\ \citenamefont
  {Apkarian}}]{yampolsky2014seeing}%
  \BibitemOpen
  \bibfield  {author} {\bibinfo {author} {\bibfnamefont {S.}~\bibnamefont
  {Yampolsky}}, \bibinfo {author} {\bibfnamefont {D.~A.}\ \bibnamefont
  {Fishman}}, \bibinfo {author} {\bibfnamefont {S.}~\bibnamefont {Dey}},
  \bibinfo {author} {\bibfnamefont {E.}~\bibnamefont {Hulkko}}, \bibinfo
  {author} {\bibfnamefont {M.}~\bibnamefont {Banik}}, \bibinfo {author}
  {\bibfnamefont {E.~O.}\ \bibnamefont {Potma}}, \ and\ \bibinfo {author}
  {\bibfnamefont {V.~A.}\ \bibnamefont {Apkarian}},\ }\href@noop {} {\bibfield
  {journal} {\bibinfo  {journal} {Nature Photonics}\ }\textbf {\bibinfo
  {volume} {8}},\ \bibinfo {pages} {650} (\bibinfo {year} {2014})}\BibitemShut
  {NoStop}%
\bibitem [{\citenamefont {Zhang}\ \emph {et~al.}(2014)\citenamefont {Zhang},
  \citenamefont {Zhen}, \citenamefont {Neumann}, \citenamefont {Day},
  \citenamefont {Nordlander},\ and\ \citenamefont {Halas}}]{zhang2014coherent}%
  \BibitemOpen
  \bibfield  {author} {\bibinfo {author} {\bibfnamefont {Y.}~\bibnamefont
  {Zhang}}, \bibinfo {author} {\bibfnamefont {Y.-R.}\ \bibnamefont {Zhen}},
  \bibinfo {author} {\bibfnamefont {O.}~\bibnamefont {Neumann}}, \bibinfo
  {author} {\bibfnamefont {J.~K.}\ \bibnamefont {Day}}, \bibinfo {author}
  {\bibfnamefont {P.}~\bibnamefont {Nordlander}}, \ and\ \bibinfo {author}
  {\bibfnamefont {N.~J.}\ \bibnamefont {Halas}},\ }\href@noop {} {\bibfield
  {journal} {\bibinfo  {journal} {Nature communications}\ }\textbf {\bibinfo
  {volume} {5}},\ \bibinfo {pages} {1} (\bibinfo {year} {2014})}\BibitemShut
  {NoStop}%
\bibitem [{\citenamefont {Zong}\ \emph {et~al.}(2019)\citenamefont {Zong},
  \citenamefont {Premasiri}, \citenamefont {Lin}, \citenamefont {Huang},
  \citenamefont {Zhang}, \citenamefont {Yang}, \citenamefont {Ren},
  \citenamefont {Ziegler},\ and\ \citenamefont {Cheng}}]{zong2019plasmon}%
  \BibitemOpen
  \bibfield  {author} {\bibinfo {author} {\bibfnamefont {C.}~\bibnamefont
  {Zong}}, \bibinfo {author} {\bibfnamefont {R.}~\bibnamefont {Premasiri}},
  \bibinfo {author} {\bibfnamefont {H.}~\bibnamefont {Lin}}, \bibinfo {author}
  {\bibfnamefont {Y.}~\bibnamefont {Huang}}, \bibinfo {author} {\bibfnamefont
  {C.}~\bibnamefont {Zhang}}, \bibinfo {author} {\bibfnamefont
  {C.}~\bibnamefont {Yang}}, \bibinfo {author} {\bibfnamefont {B.}~\bibnamefont
  {Ren}}, \bibinfo {author} {\bibfnamefont {L.~D.}\ \bibnamefont {Ziegler}}, \
  and\ \bibinfo {author} {\bibfnamefont {J.-X.}\ \bibnamefont {Cheng}},\
  }\href@noop {} {\bibfield  {journal} {\bibinfo  {journal} {Nature
  communications}\ }\textbf {\bibinfo {volume} {10}},\ \bibinfo {pages} {1}
  (\bibinfo {year} {2019})}\BibitemShut {NoStop}%
\bibitem [{\citenamefont {Keldysh}\ \emph {et~al.}(1965)\citenamefont {Keldysh}
  \emph {et~al.}}]{keldysh1965ionization}%
  \BibitemOpen
  \bibfield  {author} {\bibinfo {author} {\bibfnamefont {L.}~\bibnamefont
  {Keldysh}} \emph {et~al.},\ }\href@noop {} {\bibfield  {journal} {\bibinfo
  {journal} {Sov. Phys. JETP}\ }\textbf {\bibinfo {volume} {20}},\ \bibinfo
  {pages} {1307} (\bibinfo {year} {1965})}\BibitemShut {NoStop}%
\bibitem [{\citenamefont {Sun}\ \emph {et~al.}(1993)\citenamefont {Sun},
  \citenamefont {Vall{\'e}e}, \citenamefont {Acioli}, \citenamefont {Ippen},\
  and\ \citenamefont {Fujimoto}}]{sun1993femtosecond}%
  \BibitemOpen
  \bibfield  {author} {\bibinfo {author} {\bibfnamefont {C.-K.}\ \bibnamefont
  {Sun}}, \bibinfo {author} {\bibfnamefont {F.}~\bibnamefont {Vall{\'e}e}},
  \bibinfo {author} {\bibfnamefont {L.}~\bibnamefont {Acioli}}, \bibinfo
  {author} {\bibfnamefont {E.}~\bibnamefont {Ippen}}, \ and\ \bibinfo {author}
  {\bibfnamefont {J.}~\bibnamefont {Fujimoto}},\ }\href@noop {} {\bibfield
  {journal} {\bibinfo  {journal} {Physical Review B}\ }\textbf {\bibinfo
  {volume} {48}},\ \bibinfo {pages} {12365} (\bibinfo {year}
  {1993})}\BibitemShut {NoStop}%
\bibitem [{\citenamefont {Groeneveld}\ \emph {et~al.}(1995)\citenamefont
  {Groeneveld}, \citenamefont {Sprik},\ and\ \citenamefont
  {Lagendijk}}]{groeneveld1995femtosecond}%
  \BibitemOpen
  \bibfield  {author} {\bibinfo {author} {\bibfnamefont {R.~H.}\ \bibnamefont
  {Groeneveld}}, \bibinfo {author} {\bibfnamefont {R.}~\bibnamefont {Sprik}}, \
  and\ \bibinfo {author} {\bibfnamefont {A.}~\bibnamefont {Lagendijk}},\
  }\href@noop {} {\bibfield  {journal} {\bibinfo  {journal} {Physical Review
  B}\ }\textbf {\bibinfo {volume} {51}},\ \bibinfo {pages} {11433} (\bibinfo
  {year} {1995})}\BibitemShut {NoStop}%
\bibitem [{\citenamefont {Elsayed-Ali}\ \emph {et~al.}(1991)\citenamefont
  {Elsayed-Ali}, \citenamefont {Juhasz}, \citenamefont {Smith},\ and\
  \citenamefont {Bron}}]{elsayed1991femtosecond}%
  \BibitemOpen
  \bibfield  {author} {\bibinfo {author} {\bibfnamefont {H.}~\bibnamefont
  {Elsayed-Ali}}, \bibinfo {author} {\bibfnamefont {T.}~\bibnamefont {Juhasz}},
  \bibinfo {author} {\bibfnamefont {G.}~\bibnamefont {Smith}}, \ and\ \bibinfo
  {author} {\bibfnamefont {W.}~\bibnamefont {Bron}},\ }\href@noop {} {\bibfield
   {journal} {\bibinfo  {journal} {Physical Review B}\ }\textbf {\bibinfo
  {volume} {43}},\ \bibinfo {pages} {4488} (\bibinfo {year}
  {1991})}\BibitemShut {NoStop}%
\bibitem [{\citenamefont {Bouhelier}\ \emph {et~al.}(2005)\citenamefont
  {Bouhelier}, \citenamefont {Bachelot}, \citenamefont {Lerondel},
  \citenamefont {Kostcheev}, \citenamefont {Royer},\ and\ \citenamefont
  {Wiederrecht}}]{bouhelier2005surface}%
  \BibitemOpen
  \bibfield  {author} {\bibinfo {author} {\bibfnamefont {A.}~\bibnamefont
  {Bouhelier}}, \bibinfo {author} {\bibfnamefont {R.}~\bibnamefont {Bachelot}},
  \bibinfo {author} {\bibfnamefont {G.}~\bibnamefont {Lerondel}}, \bibinfo
  {author} {\bibfnamefont {S.}~\bibnamefont {Kostcheev}}, \bibinfo {author}
  {\bibfnamefont {P.}~\bibnamefont {Royer}}, \ and\ \bibinfo {author}
  {\bibfnamefont {G.}~\bibnamefont {Wiederrecht}},\ }\href@noop {} {\bibfield
  {journal} {\bibinfo  {journal} {Physical review letters}\ }\textbf {\bibinfo
  {volume} {95}},\ \bibinfo {pages} {267405} (\bibinfo {year}
  {2005})}\BibitemShut {NoStop}%
\bibitem [{\citenamefont {Li}\ \emph {et~al.}(2013)\citenamefont {Li},
  \citenamefont {Xiao},\ and\ \citenamefont {Zhang}}]{li2013landau}%
  \BibitemOpen
  \bibfield  {author} {\bibinfo {author} {\bibfnamefont {X.}~\bibnamefont
  {Li}}, \bibinfo {author} {\bibfnamefont {D.}~\bibnamefont {Xiao}}, \ and\
  \bibinfo {author} {\bibfnamefont {Z.}~\bibnamefont {Zhang}},\ }\href@noop {}
  {\bibfield  {journal} {\bibinfo  {journal} {New Journal of Physics}\ }\textbf
  {\bibinfo {volume} {15}},\ \bibinfo {pages} {023011} (\bibinfo {year}
  {2013})}\BibitemShut {NoStop}%
\bibitem [{\citenamefont {Brongersma}\ \emph {et~al.}(2015)\citenamefont
  {Brongersma}, \citenamefont {Halas},\ and\ \citenamefont
  {Nordlander}}]{brongersma2015plasmon}%
  \BibitemOpen
  \bibfield  {author} {\bibinfo {author} {\bibfnamefont {M.~L.}\ \bibnamefont
  {Brongersma}}, \bibinfo {author} {\bibfnamefont {N.~J.}\ \bibnamefont
  {Halas}}, \ and\ \bibinfo {author} {\bibfnamefont {P.}~\bibnamefont
  {Nordlander}},\ }\href@noop {} {\bibfield  {journal} {\bibinfo  {journal}
  {Nature nanotechnology}\ }\textbf {\bibinfo {volume} {10}},\ \bibinfo {pages}
  {25} (\bibinfo {year} {2015})}\BibitemShut {NoStop}%
\bibitem [{\citenamefont {Alabastri}\ \emph {et~al.}(2015)\citenamefont
  {Alabastri}, \citenamefont {Toma}, \citenamefont {Malerba}, \citenamefont
  {De~Angelis},\ and\ \citenamefont {Proietti~Zaccaria}}]{alabastri2015high}%
  \BibitemOpen
  \bibfield  {author} {\bibinfo {author} {\bibfnamefont {A.}~\bibnamefont
  {Alabastri}}, \bibinfo {author} {\bibfnamefont {A.}~\bibnamefont {Toma}},
  \bibinfo {author} {\bibfnamefont {M.}~\bibnamefont {Malerba}}, \bibinfo
  {author} {\bibfnamefont {F.}~\bibnamefont {De~Angelis}}, \ and\ \bibinfo
  {author} {\bibfnamefont {R.}~\bibnamefont {Proietti~Zaccaria}},\ }\href@noop
  {} {\bibfield  {journal} {\bibinfo  {journal} {ACS Photonics}\ }\textbf
  {\bibinfo {volume} {2}},\ \bibinfo {pages} {115} (\bibinfo {year}
  {2015})}\BibitemShut {NoStop}%
\bibitem [{\citenamefont {Dey}\ \emph {et~al.}(2016)\citenamefont {Dey},
  \citenamefont {Banik}, \citenamefont {Hulkko}, \citenamefont {Rodriguez},
  \citenamefont {Apkarian}, \citenamefont {Galperin},\ and\ \citenamefont
  {Nitzan}}]{dey2016observation}%
  \BibitemOpen
  \bibfield  {author} {\bibinfo {author} {\bibfnamefont {S.}~\bibnamefont
  {Dey}}, \bibinfo {author} {\bibfnamefont {M.}~\bibnamefont {Banik}}, \bibinfo
  {author} {\bibfnamefont {E.}~\bibnamefont {Hulkko}}, \bibinfo {author}
  {\bibfnamefont {K.}~\bibnamefont {Rodriguez}}, \bibinfo {author}
  {\bibfnamefont {V.}~\bibnamefont {Apkarian}}, \bibinfo {author}
  {\bibfnamefont {M.}~\bibnamefont {Galperin}}, \ and\ \bibinfo {author}
  {\bibfnamefont {A.}~\bibnamefont {Nitzan}},\ }\href@noop {} {\bibfield
  {journal} {\bibinfo  {journal} {Physical Review B}\ }\textbf {\bibinfo
  {volume} {93}},\ \bibinfo {pages} {035411} (\bibinfo {year}
  {2016})}\BibitemShut {NoStop}%
\bibitem [{\citenamefont {Crampton}\ \emph {et~al.}(2016)\citenamefont
  {Crampton}, \citenamefont {Zeytunyan}, \citenamefont {Fast}, \citenamefont
  {Ladani}, \citenamefont {Alfonso-Garcia}, \citenamefont {Banik},
  \citenamefont {Yampolsky}, \citenamefont {Fishman}, \citenamefont {Potma},\
  and\ \citenamefont {Apkarian}}]{crampton2016ultrafast}%
  \BibitemOpen
  \bibfield  {author} {\bibinfo {author} {\bibfnamefont {K.~T.}\ \bibnamefont
  {Crampton}}, \bibinfo {author} {\bibfnamefont {A.}~\bibnamefont {Zeytunyan}},
  \bibinfo {author} {\bibfnamefont {A.~S.}\ \bibnamefont {Fast}}, \bibinfo
  {author} {\bibfnamefont {F.~T.}\ \bibnamefont {Ladani}}, \bibinfo {author}
  {\bibfnamefont {A.}~\bibnamefont {Alfonso-Garcia}}, \bibinfo {author}
  {\bibfnamefont {M.}~\bibnamefont {Banik}}, \bibinfo {author} {\bibfnamefont
  {S.}~\bibnamefont {Yampolsky}}, \bibinfo {author} {\bibfnamefont {D.~A.}\
  \bibnamefont {Fishman}}, \bibinfo {author} {\bibfnamefont {E.~O.}\
  \bibnamefont {Potma}}, \ and\ \bibinfo {author} {\bibfnamefont {V.~A.}\
  \bibnamefont {Apkarian}},\ }\href@noop {} {\bibfield  {journal} {\bibinfo
  {journal} {The Journal of Physical Chemistry C}\ }\textbf {\bibinfo {volume}
  {120}},\ \bibinfo {pages} {20943} (\bibinfo {year} {2016})}\BibitemShut
  {NoStop}%
\bibitem [{\citenamefont {Caldarola}\ \emph {et~al.}(2015)\citenamefont
  {Caldarola}, \citenamefont {Albella}, \citenamefont {Cort{\'e}s},
  \citenamefont {Rahmani}, \citenamefont {Roschuk}, \citenamefont {Grinblat},
  \citenamefont {Oulton}, \citenamefont {Bragas},\ and\ \citenamefont
  {Maier}}]{caldarola2015non}%
  \BibitemOpen
  \bibfield  {author} {\bibinfo {author} {\bibfnamefont {M.}~\bibnamefont
  {Caldarola}}, \bibinfo {author} {\bibfnamefont {P.}~\bibnamefont {Albella}},
  \bibinfo {author} {\bibfnamefont {E.}~\bibnamefont {Cort{\'e}s}}, \bibinfo
  {author} {\bibfnamefont {M.}~\bibnamefont {Rahmani}}, \bibinfo {author}
  {\bibfnamefont {T.}~\bibnamefont {Roschuk}}, \bibinfo {author} {\bibfnamefont
  {G.}~\bibnamefont {Grinblat}}, \bibinfo {author} {\bibfnamefont {R.~F.}\
  \bibnamefont {Oulton}}, \bibinfo {author} {\bibfnamefont {A.~V.}\
  \bibnamefont {Bragas}}, \ and\ \bibinfo {author} {\bibfnamefont {S.~A.}\
  \bibnamefont {Maier}},\ }\href@noop {} {\bibfield  {journal} {\bibinfo
  {journal} {Nature communications}\ }\textbf {\bibinfo {volume} {6}},\
  \bibinfo {pages} {1} (\bibinfo {year} {2015})}\BibitemShut {NoStop}%
\bibitem [{\citenamefont {Albella}\ \emph {et~al.}(2014)\citenamefont
  {Albella}, \citenamefont {Alcaraz de~la Osa}, \citenamefont {Moreno},\ and\
  \citenamefont {Maier}}]{albella2014electric}%
  \BibitemOpen
  \bibfield  {author} {\bibinfo {author} {\bibfnamefont {P.}~\bibnamefont
  {Albella}}, \bibinfo {author} {\bibfnamefont {R.}~\bibnamefont {Alcaraz de~la
  Osa}}, \bibinfo {author} {\bibfnamefont {F.}~\bibnamefont {Moreno}}, \ and\
  \bibinfo {author} {\bibfnamefont {S.~A.}\ \bibnamefont {Maier}},\ }\href@noop
  {} {\bibfield  {journal} {\bibinfo  {journal} {Acs Photonics}\ }\textbf
  {\bibinfo {volume} {1}},\ \bibinfo {pages} {524} (\bibinfo {year}
  {2014})}\BibitemShut {NoStop}%
\bibitem [{\citenamefont {Albella}\ \emph {et~al.}(2013)\citenamefont
  {Albella}, \citenamefont {Poyli}, \citenamefont {Schmidt}, \citenamefont
  {Maier}, \citenamefont {Moreno}, \citenamefont {S{\'a}enz},\ and\
  \citenamefont {Aizpurua}}]{albella2013low}%
  \BibitemOpen
  \bibfield  {author} {\bibinfo {author} {\bibfnamefont {P.}~\bibnamefont
  {Albella}}, \bibinfo {author} {\bibfnamefont {M.~A.}\ \bibnamefont {Poyli}},
  \bibinfo {author} {\bibfnamefont {M.~K.}\ \bibnamefont {Schmidt}}, \bibinfo
  {author} {\bibfnamefont {S.~A.}\ \bibnamefont {Maier}}, \bibinfo {author}
  {\bibfnamefont {F.}~\bibnamefont {Moreno}}, \bibinfo {author} {\bibfnamefont
  {J.~J.}\ \bibnamefont {S{\'a}enz}}, \ and\ \bibinfo {author} {\bibfnamefont
  {J.}~\bibnamefont {Aizpurua}},\ }\href@noop {} {\bibfield  {journal}
  {\bibinfo  {journal} {The Journal of Physical Chemistry C}\ }\textbf
  {\bibinfo {volume} {117}},\ \bibinfo {pages} {13573} (\bibinfo {year}
  {2013})}\BibitemShut {NoStop}%
\bibitem [{\citenamefont {Alessandri}\ and\ \citenamefont
  {Lombardi}(2016)}]{alessandri2016enhanced}%
  \BibitemOpen
  \bibfield  {author} {\bibinfo {author} {\bibfnamefont {I.}~\bibnamefont
  {Alessandri}}\ and\ \bibinfo {author} {\bibfnamefont {J.~R.}\ \bibnamefont
  {Lombardi}},\ }\href@noop {} {\bibfield  {journal} {\bibinfo  {journal}
  {Chemical reviews}\ }\textbf {\bibinfo {volume} {116}},\ \bibinfo {pages}
  {14921} (\bibinfo {year} {2016})}\BibitemShut {NoStop}%
\bibitem [{\citenamefont {Barreda}\ \emph {et~al.}(2019)\citenamefont
  {Barreda}, \citenamefont {Saiz}, \citenamefont {Gonz{\'a}lez}, \citenamefont
  {Moreno},\ and\ \citenamefont {Albella}}]{barreda2019recent}%
  \BibitemOpen
  \bibfield  {author} {\bibinfo {author} {\bibfnamefont {A.}~\bibnamefont
  {Barreda}}, \bibinfo {author} {\bibfnamefont {J.}~\bibnamefont {Saiz}},
  \bibinfo {author} {\bibfnamefont {F.}~\bibnamefont {Gonz{\'a}lez}}, \bibinfo
  {author} {\bibfnamefont {F.}~\bibnamefont {Moreno}}, \ and\ \bibinfo {author}
  {\bibfnamefont {P.}~\bibnamefont {Albella}},\ }\href@noop {} {\bibfield
  {journal} {\bibinfo  {journal} {AIP Advances}\ }\textbf {\bibinfo {volume}
  {9}},\ \bibinfo {pages} {040701} (\bibinfo {year} {2019})}\BibitemShut
  {NoStop}%
\bibitem [{\citenamefont {Bontempi}\ \emph {et~al.}(2014)\citenamefont
  {Bontempi}, \citenamefont {Salmistraro}, \citenamefont {Ferroni},
  \citenamefont {Depero},\ and\ \citenamefont
  {Alessandri}}]{bontempi2014probing}%
  \BibitemOpen
  \bibfield  {author} {\bibinfo {author} {\bibfnamefont {N.}~\bibnamefont
  {Bontempi}}, \bibinfo {author} {\bibfnamefont {M.}~\bibnamefont
  {Salmistraro}}, \bibinfo {author} {\bibfnamefont {M.}~\bibnamefont
  {Ferroni}}, \bibinfo {author} {\bibfnamefont {L.~E.}\ \bibnamefont {Depero}},
  \ and\ \bibinfo {author} {\bibfnamefont {I.}~\bibnamefont {Alessandri}},\
  }\href@noop {} {\bibfield  {journal} {\bibinfo  {journal} {Nanotechnology}\
  }\textbf {\bibinfo {volume} {25}},\ \bibinfo {pages} {465705} (\bibinfo
  {year} {2014})}\BibitemShut {NoStop}%
\bibitem [{\citenamefont {Wang}\ \emph {et~al.}(2011)\citenamefont {Wang},
  \citenamefont {Shi}, \citenamefont {She},\ and\ \citenamefont
  {Mu}}]{wang2011using}%
  \BibitemOpen
  \bibfield  {author} {\bibinfo {author} {\bibfnamefont {X.}~\bibnamefont
  {Wang}}, \bibinfo {author} {\bibfnamefont {W.}~\bibnamefont {Shi}}, \bibinfo
  {author} {\bibfnamefont {G.}~\bibnamefont {She}}, \ and\ \bibinfo {author}
  {\bibfnamefont {L.}~\bibnamefont {Mu}},\ }\href@noop {} {\bibfield  {journal}
  {\bibinfo  {journal} {Journal of the American Chemical Society}\ }\textbf
  {\bibinfo {volume} {133}},\ \bibinfo {pages} {16518} (\bibinfo {year}
  {2011})}\BibitemShut {NoStop}%
\bibitem [{\citenamefont {Khorasaninejad}\ \emph
  {et~al.}(2012{\natexlab{a}})\citenamefont {Khorasaninejad}, \citenamefont
  {Dhindsa}, \citenamefont {Walia}, \citenamefont {Patchett},\ and\
  \citenamefont {Saini}}]{khorasaninejad2012highly}%
  \BibitemOpen
  \bibfield  {author} {\bibinfo {author} {\bibfnamefont {M.}~\bibnamefont
  {Khorasaninejad}}, \bibinfo {author} {\bibfnamefont {N.}~\bibnamefont
  {Dhindsa}}, \bibinfo {author} {\bibfnamefont {J.}~\bibnamefont {Walia}},
  \bibinfo {author} {\bibfnamefont {S.}~\bibnamefont {Patchett}}, \ and\
  \bibinfo {author} {\bibfnamefont {S.}~\bibnamefont {Saini}},\ }\href@noop {}
  {\bibfield  {journal} {\bibinfo  {journal} {Applied Physics Letters}\
  }\textbf {\bibinfo {volume} {101}},\ \bibinfo {pages} {173114} (\bibinfo
  {year} {2012}{\natexlab{a}})}\BibitemShut {NoStop}%
\bibitem [{\citenamefont {Khorasaninejad}\ \emph
  {et~al.}(2012{\natexlab{b}})\citenamefont {Khorasaninejad}, \citenamefont
  {Walia},\ and\ \citenamefont {Saini}}]{khorasaninejad2012enhanced}%
  \BibitemOpen
  \bibfield  {author} {\bibinfo {author} {\bibfnamefont {M.}~\bibnamefont
  {Khorasaninejad}}, \bibinfo {author} {\bibfnamefont {J.}~\bibnamefont
  {Walia}}, \ and\ \bibinfo {author} {\bibfnamefont {S.}~\bibnamefont
  {Saini}},\ }\href@noop {} {\bibfield  {journal} {\bibinfo  {journal}
  {Nanotechnology}\ }\textbf {\bibinfo {volume} {23}},\ \bibinfo {pages}
  {275706} (\bibinfo {year} {2012}{\natexlab{b}})}\BibitemShut {NoStop}%
\bibitem [{\citenamefont {Huang}\ \emph {et~al.}(2011)\citenamefont {Huang},
  \citenamefont {Zhao}, \citenamefont {Zhang}, \citenamefont {Luo},
  \citenamefont {Liu}, \citenamefont {Zapien}, \citenamefont {Surya},\ and\
  \citenamefont {Lee}}]{huang2011enhanced}%
  \BibitemOpen
  \bibfield  {author} {\bibinfo {author} {\bibfnamefont {J.-A.}\ \bibnamefont
  {Huang}}, \bibinfo {author} {\bibfnamefont {Y.-Q.}\ \bibnamefont {Zhao}},
  \bibinfo {author} {\bibfnamefont {X.-J.}\ \bibnamefont {Zhang}}, \bibinfo
  {author} {\bibfnamefont {L.-B.}\ \bibnamefont {Luo}}, \bibinfo {author}
  {\bibfnamefont {Y.-K.}\ \bibnamefont {Liu}}, \bibinfo {author} {\bibfnamefont
  {J.~A.}\ \bibnamefont {Zapien}}, \bibinfo {author} {\bibfnamefont
  {C.}~\bibnamefont {Surya}}, \ and\ \bibinfo {author} {\bibfnamefont {S.-T.}\
  \bibnamefont {Lee}},\ }\href@noop {} {\bibfield  {journal} {\bibinfo
  {journal} {Applied Physics Letters}\ }\textbf {\bibinfo {volume} {98}},\
  \bibinfo {pages} {183108} (\bibinfo {year} {2011})}\BibitemShut {NoStop}%
\bibitem [{\citenamefont {Abedin}\ \emph {et~al.}(2022)\citenamefont {Abedin},
  \citenamefont {Roy}, \citenamefont {Jin}, \citenamefont {Xia}, \citenamefont
  {Brueck},\ and\ \citenamefont {Potma}}]{abedin2022}%
  \BibitemOpen
  \bibfield  {author} {\bibinfo {author} {\bibfnamefont {S.}~\bibnamefont
  {Abedin}}, \bibinfo {author} {\bibfnamefont {K.}~\bibnamefont {Roy}},
  \bibinfo {author} {\bibfnamefont {X.}~\bibnamefont {Jin}}, \bibinfo {author}
  {\bibfnamefont {H.}~\bibnamefont {Xia}}, \bibinfo {author} {\bibfnamefont
  {S.~R.~J.}\ \bibnamefont {Brueck}}, \ and\ \bibinfo {author} {\bibfnamefont
  {E.~O.}\ \bibnamefont {Potma}},\ }\href {\doibase 10.1021/acs.jpcc.2c01642}
  {\bibfield  {journal} {\bibinfo  {journal} {Journal of Physical Chemistry C}\
  } (\bibinfo {year} {2022}),\ 10.1021/acs.jpcc.2c01642}\BibitemShut {NoStop}%
\bibitem [{\citenamefont {Edward}\ and\ \citenamefont
  {Palik}(1985)}]{edward1985handbook}%
  \BibitemOpen
  \bibfield  {author} {\bibinfo {author} {\bibfnamefont {D.~P.}\ \bibnamefont
  {Edward}}\ and\ \bibinfo {author} {\bibfnamefont {I.}~\bibnamefont {Palik}},\
  }\href@noop {} {\enquote {\bibinfo {title} {Handbook of optical constants of
  solids},}\ } (\bibinfo {year} {1985})\BibitemShut {NoStop}%
\bibitem [{\citenamefont {Benassi}\ and\ \citenamefont
  {Fan}(2021)}]{Benassi2021}%
  \BibitemOpen
  \bibfield  {author} {\bibinfo {author} {\bibfnamefont {E.}~\bibnamefont
  {Benassi}}\ and\ \bibinfo {author} {\bibfnamefont {H.}~\bibnamefont {Fan}},\
  }\href {\doibase https://doi.org/10.1016/j.saa.2020.119026} {\bibfield
  {journal} {\bibinfo  {journal} {Spectrochimica Acta Part A: Molecular and
  Biomolecular Spectroscopy}\ }\textbf {\bibinfo {volume} {246}},\ \bibinfo
  {pages} {119026} (\bibinfo {year} {2021})}\BibitemShut {NoStop}%
\bibitem [{\citenamefont {Perry}\ \emph {et~al.}(1997)\citenamefont {Perry},
  \citenamefont {Green},\ and\ \citenamefont {Maloney}}]{perry1997perry}%
  \BibitemOpen
  \bibfield  {author} {\bibinfo {author} {\bibfnamefont {R.~H.}\ \bibnamefont
  {Perry}}, \bibinfo {author} {\bibfnamefont {D.~W.}\ \bibnamefont {Green}}, \
  and\ \bibinfo {author} {\bibfnamefont {J.}~\bibnamefont {Maloney}},\
  }\href@noop {} {\bibfield  {journal} {\bibinfo  {journal} {Seventh,
  International edition}\ } (\bibinfo {year} {1997})}\BibitemShut {NoStop}%
\end{thebibliography}%

\end{document}